\RequirePackage{lineno}
\documentclass[aps,prc,twocolumn, floatfix]{revtex4}
\usepackage{graphicx}
\usepackage{dcolumn}
\usepackage{bm}

\begin{document}

%\setpagewiselinenumbers
%\modulolinenumbers[5]
%\linenumbers

\title{A method for probing the formation of quark matter}

\author{Gao-Chan Yong$^{1,2}$}
%\email[]{yonggaochan@impcas.ac.cn}

\affiliation{
$^1$Institute of Modern Physics, Chinese Academy of Sciences, Lanzhou 730000, China\\
$^2$School of Nuclear Science and Technology, University of Chinese Academy of Sciences, Beijing 100049, China
}

\begin{abstract}

Based on a multi-phase transport model for relativistic heavy-ion collisions, effects of the parton scatterings on the production of strangeness in relativistic heavy-ion collisions are studied.
It is found that the distributions of strange quark and strange baryon, especially for the double strangeness $\Xi^{-}$, are significantly affected by the parton scatterings in heavy-ion collisions below $\sqrt{s_{NN}}\sim$ 10 GeV. Given parton scatterings as a signal of the formation of quark matter, the transverse momentum distribution of the ratio of single and double strangeness $(\Lambda+\Sigma^{0})/\Xi^{-}$ produced in heavy-ion collisions may serve as a potential probe of the emergence of quark matter, or equivalently, the occurrence of hadron-quark phase transition in relativistic heavy-ion collisions.

\end{abstract}

\maketitle

%\section{Introduction}
%
Understanding the properties of nuclear matter under conditions of extreme energy and baryon density, or equivalently, exploring the phase structure of hot and dense nuclear matters, is one of the main goals of relativistic heavy-ion collisions \cite{pr1,pr2,pr3}. Due to continuous efforts of many people, a lot of circumstantial signals in relativistic heavy-ion collisions have been observed for the formation of quark-gluon plasma (QGP) \cite{qgp1,qgp2}. It is general considered that at small baryon chemical potential and high temperature, the transition from hadronic phase to QGP phase is a smooth crossover, whereas a first order phase transition is expected at high baryon chemical potential region \cite{pr3}.
Since at large baryon chemical potential, theoretical calculations are still struggling with difficulties \cite{lqcd1,lqcd2}, relativistic heavy-ion collisions' experimental program provides a unique opportunity to explore the QCD phase structure that is not yet accessible to theoretical calculations.

The study of the QCD phase diagram of strongly interacting nuclear matter is the current focus of many research activities worldwide, both theoretically and experimentally \cite{pt1,pt2,ptko22,guo2021,nara18}. Indeed, mapping the QCD phase diagram is the major scientific goal of the beam energy scan (BES) program in heavy-ion collisions \cite{pr2,bes19,besa,besb}.
Also the Compressed Baryonic Matter (CBM) experiment aims to study the Equation of State (EoS) of dense
baryonic matter, a possible first order phase transition as well as the existence of the critical end point (CEP) in the baryon rich domain by measuring rare probes \cite{CBM17,CBM21}.
Nuclotron-based Ion Collider fAcility (NICA) allows to study the EoS of dense baryonic matter and the QCD phase transition by measuring multi-strange hyperons and hypernuclei with the Multi-Purpose Detector (MPD) \cite{NICA19}.
To this end, strange mesons or baryons have been suggested to identify the softness of dense nuclear matter, the phase boundary and onset of deconfinement \cite{raf82,adam20,cas2021,chen2020}. The constructed or being constructed many other facilities worldwide all have related research projects/plans \cite{qm2018}. On the other hand, whether there is hadron-quark phase transition in neutron-stars (NSs) with central densities of several times nuclear saturation density, is also of great interest in the study of neutron star structure \cite{akm1998,nature2020,liapj2020,hu2021,kojo2021,ran2021,xie2021} and gravitational-wave (GW) emission \cite{gw2019,gw2018, gw20182}. The study of the phase transition of QCD matter from earth to heaven is thought to have crucial implications toward an unprecedented understanding of the early and present universe \cite{ann2006}.

So far, both the single strangeness $\Lambda$ and the double strangeness $\Xi^{-}$ have been measured by E895, NA49, HADES and STAR experimental collaborations in heavy-ion collisions with various colliding energies \cite{chung2001,exp2003,thesis2004,exp2008,exp2009,exp2015,adam20,exp2021}, while their connections to the softness of dense nuclear matter, the phase boundary and onset of deconfinement are not sufficiently studied. Very recently, it is shown in Refs.~\cite{cas2021,yongrcas2022} that the doubly strange $\Xi$ hyperon is expected to be more sensitive to the stiffness of the nuclear EoS at high densities, simply because the doubly strange $\Xi$ hyperon is more likely to be produced at maximum compression of nucleus-nucleus collision than the single strangeness. Due to strangeness conservation, once produced, the strange particles are rarely absorbed by surrounding matter. To probe the phase-transition boundary of QCD matter, it is thus of great interest to see if the double strangeness $\Xi$ can be used to explore the phase-transition boundary of QCD matter, or equivalently, the emergence of quark matter. To reduce systematic errors, besides the double strangeness $\Xi^{-}$, the singly strange hyperons $\Lambda+\Sigma^{0}$ are also used as counterparts. As the study of the nonmonotonic energy dependence of net-proton number fluctuations $\kappa\sigma^{2}$ indicates that the phase-transition of QCD matter from hadronic phase to QGP phase very likely occurs in the heavy-ion collisions at $\sqrt{s_{NN}}<$ 10 GeV \cite{bj1,new2021}, in the present study, the collision energies are concentrated in the range of $\sqrt{s_{NN}}<$ 10 GeV.

%\section{The AMPT model with different modes}
%
To match the present study of probing the emergence of quark matter, or equivalently, the phase-transition boundary of QCD matter from hadronic phase to QGP phase, a multi-phase transport (AMPT) model \cite{AMPT2005} is recently extended so that it can perform not only multi-phase transport simulations with both parton and hadron degrees of freedom but also pure hadron cascade with hadronic mean-field potentials. Moreover, reaction channels relating to the $\Xi$ production have been replenished \cite{cas2021}.

As a Monte Carlo parton and hadron transport model, the AMPT model consists of four components, i.e., a fluctuating initial condition, partonic interactions, conversion from the partonic to the hadronic matter, and hadronic interactions \cite{AMPT2005}. The model has been extensively applied to heavy-ion collisions at RHIC and LHC energies \cite{nst2021}. In the AMPT model, $\pi$, $\rho$, $\omega$, $\eta$, $K$, $K^*$, $\phi$, $N$, $\Delta$, $N^*(1440)$, $N^*(1535)$, $\Lambda$, $\Sigma$, $\Xi$ and $\Omega$ are included \cite{deu2009}. In the used string melting AMPT model (AMPT-SM), the initial partons are produced through the intermediate step of decomposition of hardrons formed via Lund string fragmentation as in the
HIJING model \cite{wang1,wang2}. The original Lund string fragmentation parameters $a$ = 0.55, $b$ = 0.15/GeV$^{2}$, the strong coupling constant $\alpha_{s}$= 0.33, the parton cross section $\sigma$ = 3 mb are used \cite{linab14}. Scatterings of melted partons are described by the Zhang's Parton Cascade (ZPC) model \cite{zhang1}. After a quark coalescence model is used for hadronization, subsequent hadronic  interactions are described by a hadronic cascade based on a relativistic transport (ART) model \cite{art}.

To study heavy-ion collisions at low energies, the present AMPT-SM model includes the effects of finite nuclear thickness \cite{thick3,thick1,thick2}. This approach actually restores a colliding nucleus from a disk shape to an elliptical shape, the details can be found in Ref.~\cite{thick3}. Also a pure hadron cascade model (AMPT-HC) is recently extended \cite{cas2021}. In the AMPT-HC model, the Woods-Saxon nucleon density distribution and local Thomas-Fermi approximation are used to initialize the position and momentum of each nucleon in colliding projectile and target. The parton degree of freedom is switched off. In addition to the usual elastic and inelastic collisions, hadron potentials with the test-particle method are applied to nucleons, baryonic resonances, strangenesses as well as their antiparticles \cite{cas2021,yongrcas2022}.

%\section{Results and Discussions}
%
\begin{figure}[t]
\centering
\vspace{0.0cm}
\includegraphics[width=0.42\textwidth]{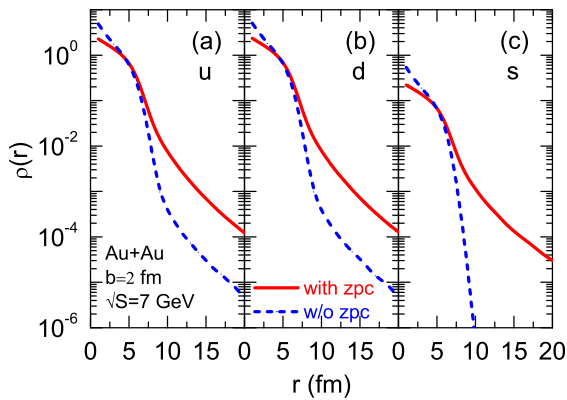}
\caption{Quark density distribution as a function of distance $r$ in the center of mass framework before hadronization in the central Au+Au collisions at $\sqrt{s_{NN}}$ = 7 GeV with and without parton cascade. Panels (a), (b) and (c) show distributions of nonstrange quarks $u$+$\bar{u}$ and $d$+$\bar{d}$ as well as strange quarks $s$+$\bar{s}$, respectivley.} \label{quark}
\end{figure}
\begin{figure}[t]
\centering
\vspace{-0.5cm}
\includegraphics[width=0.5\textwidth]{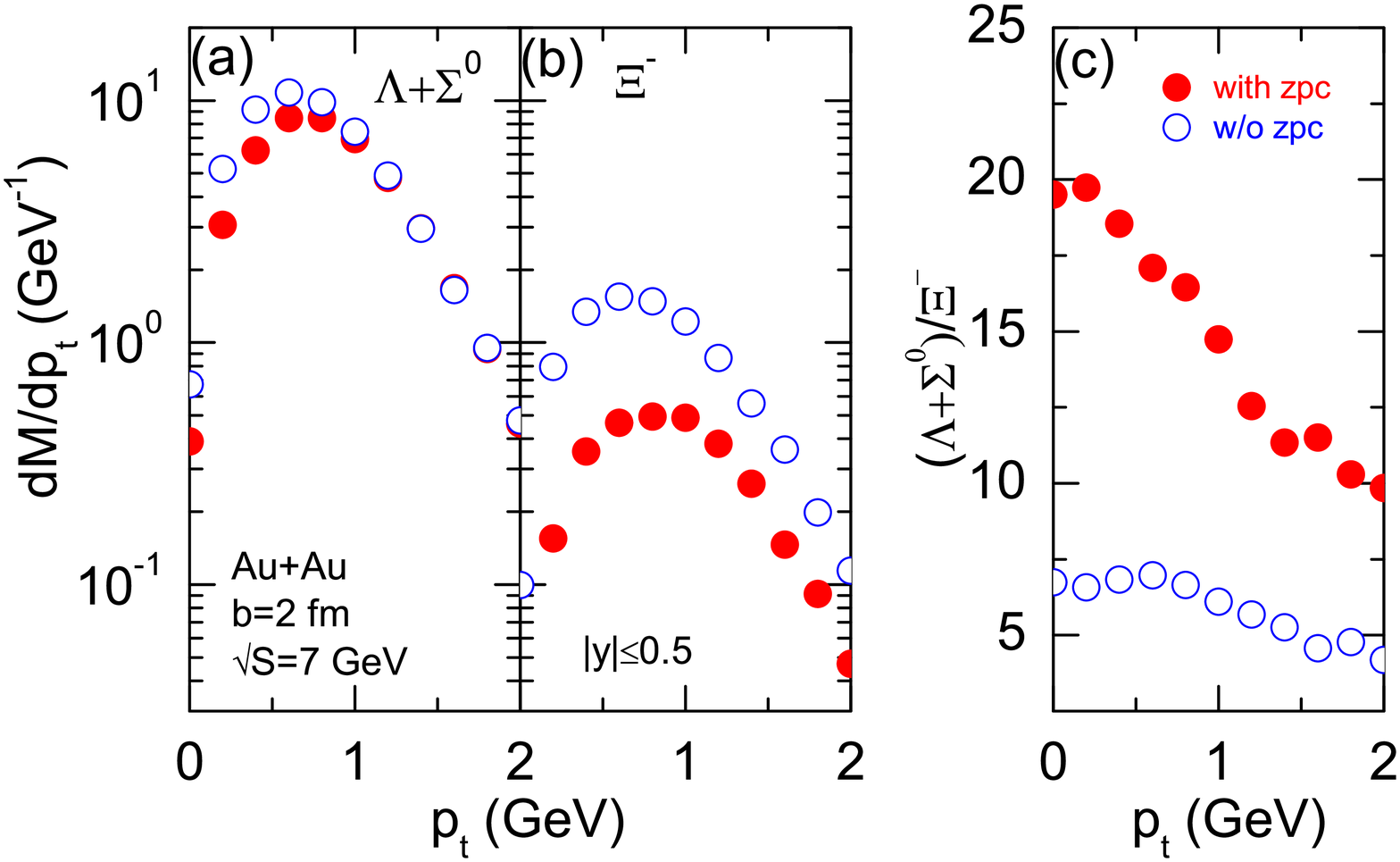}
\caption{Effects of the parton cascade on the transverse momentum distribution of mid-rapidity $\Lambda+\Sigma^{0}$ and $\Xi^{-}$ productions and their yields ratio given by the AMPT-SM model with or without parton cascade ZPC in the central Au+Au collisions at $\sqrt{s_{NN}}$ = 7 GeV.} \label{sdis1}
\end{figure}
Comparing with the HIJING model \cite{wang1,wang2} and the default AMPT model \cite{AMPT2005}, the most prominent characteristic of the AMPT-SM model is its \emph{free} parton production thus the parton cascade always works. To show the effects of parton cascade on the quark density distribution at quark freeze-out stage in heavy-ion collisions at lower energies, quark density distribution as a function of distance $r$ in the center of mass framework before hadronization in the central Au+Au collisions at $\sqrt{s_{NN}}$ = 7 GeV is demonstrated in Figure~\ref{quark}. Note here that the quark density distribution shown here is the quarks at freeze-out stage, i.e., after their propagation, while not their initial formation via constituent quarks as done in AMPT-SM model.
It is seen that with parton cascade the freezed-out quark has sparser distribution compared with that without parton cascade. It is interesting to see that, comparing panels (a), (b) with panel (c), the strange quark density distribution is much affected by the parton cascade although the yield of the strange quark is overall smaller than that of nonstrange quark. This is because in central heavy-ion collisions the strangeness is more preferentially produced in the center of compressed dense matter \cite{cas2021}. More compacted distribution of the strangeness is thus more easily scattered away or dispersed through the parton cascade. Therefore one sees in Figure~\ref{quark} the strange quark distribution is much affected by the proton cascade.

After parton scatterings, a quark coalescence model is used to describe the hadronization process. It combines a quark with a nearest antiquark to form a meson and combines three nearest quarks (or antiquarks) into a baryon (or an antibaryon), regardless of the relative momentum among the coalescing partons. The parton cascade ZPC affects partonic spatial distribution in various directions through parton scatterings and propagation. In practice, the relative momentum cuts $\Delta p= 3,2,1$ GeV are operated in Au+Au at $\sqrt{s_{NN}}$ = 4 GeV. It is found that both the yields of strangeness and their ratios are less sensitive to such momentum cuts while quarks coalesce in hadronization process. Although the above quark coalescence model includes almost all the formations of mesons and baryons listed in the HIJING program \cite{wang2}, here mainly the singly strange $\Lambda+\Sigma^{0}$ and the doubly strange $\Xi^{-}$ hyperons are analyzed due to their peculiar sensitivity to emergence of quark matter, or equivalently, the phase transition of QCD matter.

Figure~\ref{sdis1} shows the yields of $\Lambda+\Sigma^{0}$ and $\Xi^{-}$ and their ratio in the central Au+Au collisions at $\sqrt{s_{NN}}$ = 7 GeV with and without parton scatterings. Since with the parton cascade quark has sparser distribution especially for strange quark as shown in Figure~\ref{quark}, one sees both $\Lambda+\Sigma^{0}$ and $\Xi^{-}$ are less produced with parton scatterings. Because the doubly strange $\Xi^{-}$ hyperon is the coalescence of three nearest quarks including two strange quarks, its multiplicity is more affected by the parton cascade. From panel (c) of Figure~\ref{sdis1}, one can see that the yields ratio of $\Lambda+\Sigma^{0}$ and $\Xi^{-}$ is very sensitive to the parton cascade.

\begin{figure}[t]
\centering
\includegraphics[width=0.42\textwidth]{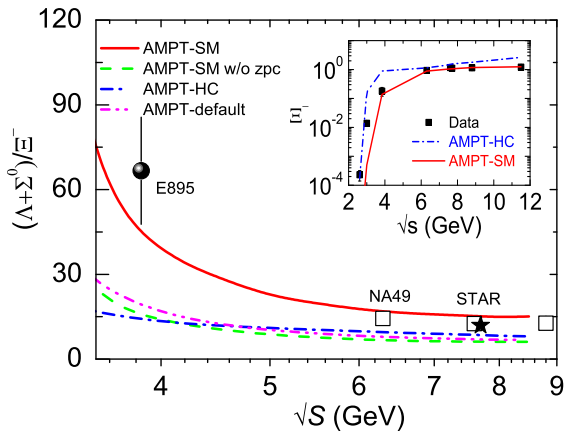}
\caption{The ratio of mid-rapidity ($\Lambda+\Sigma^{0}$)/$\Xi^{-}$ as a function of $\sqrt{s_{NN}}$ given by various modes of the AMPT model in heavy-ion collisions. The inset shows the $\Xi^{-}$ production as a function of $\sqrt{s_{NN}}$ with the AMPT-HC and the AMPT-SM (the first three data points have no rapidity cut while others have mid-rapidity cut $|y|\leq0.5$). } \label{plus}
\end{figure}
\begin{figure}[t]
\centering
\vspace{-0.5cm}
\includegraphics[width=0.5\textwidth]{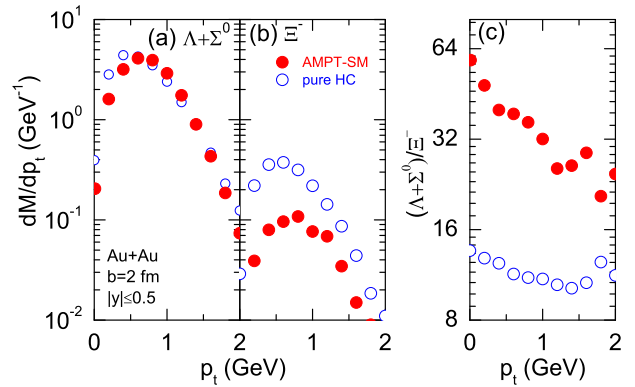}
\caption{Transverse momentum distribution of mid-rapidity $\Lambda+\Sigma^{0}$ and $\Xi^{-}$ productions and their yields ratio given by the transport model including parton cascade (AMPT-SM) and pure hadron transport model (AMPT-HC) in the central Au+Au collisions at $\sqrt{s_{NN}}$ = 4 GeV.} \label{sdis2}
\end{figure}
In fact, the essential difference of the pure hadron transport model and the multi-phase transport model ( AMPT-HC vs AMPT-SM) is whether quarks are confined or not. With this consideration, the AMPT-SM model without parton cascade should be similar to the pure hadron transport model AMPT-HC to a great extent and
my actual computations support this expectation. Figure~\ref{plus} shows the ratios of $(\Lambda+\Sigma^{0})/\Xi^{-}$ as a function of $\sqrt{s_{NN}}$. It is seen that the predicted ratios have large differences between the AMPT-SM mode with free quark transport and those modes without free quarks (including the AMPT-SM without ZPC, the AMPT-HC as well as the AMPT-default).
Figure~\ref{sdis2} shows different consequences on the $\Lambda+\Sigma^{0}$ and $\Xi^{-}$ productions as well as their ratio with the multi and single-phase transport models (i.e., AMPT-SM and AMPT-HC). Compared with that shown in Figure~\ref{sdis1}, one again sees similar behavior of the transverse momentum distributions of the $\Lambda+\Sigma^{0}$ and $\Xi^{-}$ as well as their yields ratio. To think more deeply, if a nucleus-nucleus colliding process can only be described by a transport model including parton cascade, there must appear free quarks or quark matter, i.e., there must be occurrence of phase transition from hadronic matter to quark matter. Although the EoS of nuclear matter affects the strangeness production~\cite{cas2021,yongrcas2022}, its effects on the $(\Lambda+\Sigma^{0})/\Xi^{-}$ are less than one fifth that of parton cascade studied here. Hence the variation of the EoS of nuclear matter does not evidently change the conclusion drawn here.

I actually compared the simulated results of the ratios of $(\Lambda+\Sigma^{0})/\Xi^{-}$ and the yields of $\Xi^{-}$ at various beam energies by using the AMPT-SM and the AMPT-HC mode, respectively, with related experimental data in the literature \cite{chung2001,exp2003,thesis2004,exp2008,exp2009,exp2015,adam20,exp2021} as shown in Figure~\ref{plus}.
It is found that above $\sqrt{s_{NN}}$ $\simeq$ 4 GeV, the AMPT-SM's results roughly fit the data well, thus indicating the occurrence of quark matter in Au+Au collisions above $\sqrt{s_{NN}}$ $\simeq$ 4 GeV. However, a definite conclusion could not be drawn before carrying out more detailed studies on the model dependence.

\begin{figure}[t]
\centering
\vspace{0.0cm}
\includegraphics[width=0.45\textwidth]{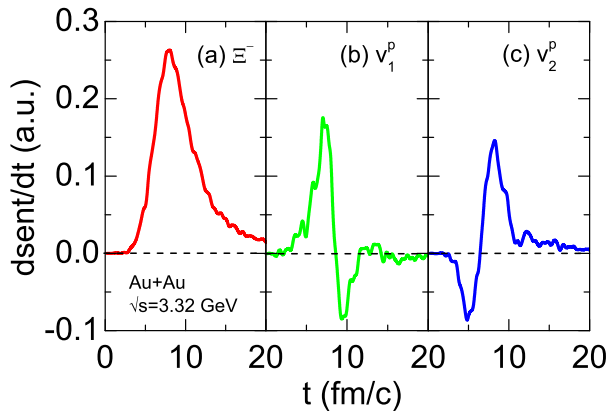}
\caption{Sensitivity increments of strangeness $\Xi^{-}$ production (a), proton sideward flow (slope of $v1=<p_{x}/p_{t}>$) (b) and elliptic flow (integration of $v2=<\frac{p_{x}^{2}-p_{y}^{2}}{p_{x}^{2}+p_{y}^{2}}>$ at mid-rapidity $|y|\leq0.2$) (c) to the EoS as a function of time, simulated by the AMPT-HC model.} \label{senst}
\end{figure}
To search for the softest point (associated with the transition from hadronic matter to quark-gluon plasma) of the EoS in QCD matter, the excitation functions of proton elliptic flow and sideward flow have been extensively studied in the literature based on the experimental measurements at Alternating Gradient Synchrotron (AGS) at the Brookhaven National Laboratory \cite{ags1,ags2}, but unfortunately no definite conclusion was made on whether the hadron-quark phase transition occurs or not in Au+Au collisions at AGS energy regime. It is necessary to make comparative studies so as to find out the advantages and disadvantages among various observables while probing the formation of quark matter in heavy-ion collisions. Figure~\ref{senst} shows sensitivity increments of strangeness $\Xi^{-}$ production, proton sideward flow and elliptic flow to the EoS as a function of time. By definition, the variable $sent = V(t)_{EoS1}-V(t)_{EoS2}$ with $V(t)_{EoS1,2}$ being two different values of an observable at time point $t$ with two different EoSs. The sensitivity increment is $d sent/dt$, which reflecting the situation when and where the EoS plays a role during the evolution of an observable in heavy-ion collisions. Due to strangeness's minor interactions with surrounding matter once produced, it is seen that sensitivity of the strangeness $\Xi^{-}$ production to the EoS mainly occurs at certain phase space of dense matter while the sensitivity of proton flow to the EoS exhibits complexity and there are cancellations among sensitivity increments at different phase space points. Comparison shows that strangeness production may be more suitable to probe the properties of dense nuclear matter, especially for the possibly localized hadron-quark phase transition in heavy-ion collisions.

%\section{Conclusions}
%
In summary, an investigation on the signal of the emergence of quark matter formed in heavy-ion collisions is carried out. It is shown that the distributions of strange quark and strange baryon, especially for the double strangeness $\Xi^{-}$, are quite sensitive to the parton scatterings in heavy-ion collisions below $\sqrt{s_{NN}}\sim$ 10 GeV. Given parton scatterings as a signal of quark matter formed in heavy-ion collisions, the $(\Lambda+\Sigma^{0})/\Xi^{-}$ ratio could be a potential probe of the occurrence of quark matter in relativistic heavy-ion collisions.

%\section{Acknowledgments}
%
The author thanks Zi-Wei Lin for providing the algorithm on the finite nuclear thickness of AMPT model. This work is supported by the National Natural Science Foundation of China under Grant No. 12275322 and the Strategic Priority Research Program of Chinese Academy of Sciences with Grant No. XDB34030000.

\end{document}